\documentclass[aps,prl,superscriptaddress,preprint]{revtex4}

\usepackage[dvips]{graphicx}
\usepackage{indentfirst}
\usepackage{bm}
\usepackage{enumerate}

\def\be{\begin{equation}}
\def\ee{\end{equation}}
\def\bea{\begin{eqnarray}}
\def\eea{\end{eqnarray}}
\newcommand{\ket}[1]{\mbox{$|#1\rangle$}}
\newcommand{\bra}[1]{\mbox{$\langle#1|$}}
\newcommand{\avg}[1]{\mbox{$\langle#1\rangle$}}

\def\bfp0{{\bf{p_0}}}

\def\Aeff{A_{\footnotesize\textrm{eff}}}
\newcommand{\opdagger}[2]{\mbox{$\hat{#1}_{#2}^{\dagger}$}}
\newcommand{\op}[2]{\mbox{$\hat{#1}_{#2}$}}
\def\Gammapl{\Gamma_{\footnotesize\textrm{pl}}}

\begin{document}
\title{A single-photon transistor using nano-scale surface plasmons}

\author{D.E. Chang}
\affiliation{Physics Department, Harvard University, Cambridge, MA
02138}

\author{A.S. S{\o}rensen}
\affiliation{QUANTOP, Danish Quantum Optics Center and Niels Bohr
Institute, DK-2100 Copenhagen \O, Denmark}

\author{E.A. Demler}
\affiliation{Physics Department, Harvard University, Cambridge, MA
02138}

\author{M.D. Lukin}
\affiliation{Physics Department, Harvard University, Cambridge, MA
02138}

\date{\today}

\begin{abstract}
It is well known that light quanta~(photons) can interact with
each other in nonlinear media, much like massive particles do, but
in practice these interactions are usually very weak.  Here we
describe a novel approach to realize strong nonlinear interactions
at the single-photon level. Our method makes use of recently
demonstrated efficient coupling between individual optical
emitters and tightly confined, propagating surface plasmon
excitations on conducting nanowires. We show that this system can
act as a nonlinear two-photon switch for incident photons
propagating along the nanowire, which can be coherently controlled
using quantum optical techniques.  As a novel application, we
discuss how the interaction can be tailored to create a
single-photon transistor, where the presence or absence of a
single incident photon in a ``gate'' field is sufficient to
completely control the propagation of subsequent ``signal''
photons.

\end{abstract}

\maketitle

In analogy with the electronic transistor, a photonic transistor
is a device where a small optical ``gate'' field is used to
control the propagation of another optical ``signal'' field via a
nonlinear optical interaction~\cite{boyd92,gibbs85}. Its
fundamental limit is the single-photon transistor, where the
propagation of the signal field is controlled by the presence or
absence of a single photon in the gate field. Nonlinear devices of
this kind would have a number of interesting applications ranging
from optical communication and computation~\cite{gibbs85} to
quantum information processing~\cite{bouwmeester00}. However,
their practical realization is challenging because the requisite
single-photon nonlinearities are generally very
weak~\cite{boyd92}. While several schemes for producing
nonlinearities at the single-photon level are currently being
explored, ranging from resonantly enhanced nonlinearities of
atomic
ensembles~\cite{schmidt96,harris97,harris98,lukin03,fleischhauer05}
to individual atoms strongly coupled to photons in cavity quantum
electrodynamics~(QED)~\cite{vahala04,miller05,duan04,birnbaum05,waks06},
a robust, practical approach has yet to emerge.

Recently, a new method to achieve strong coupling between light
and matter has been proposed~\cite{chang06a} and experimentally
demonstrated~\cite{akimov07}. It makes use of the tight
concentration of optical fields associated with guided surface
plasmons (SPs) on conducting nanowires to achieve strong
interaction with individual optical emitters. In essence, the
tight localization of these fields causes the nanowire to act as a
very efficient lens that directs the majority of the spontaneously
emitted light into the SP modes, resulting in efficient generation
of single plasmons~(single photons)~\cite{chang06a}. While this
process is essentially a linear optical effect, as it only
involves one photon at a time, here we show that such a system
also allows for the realization of remarkable nonlinear optical
phenomena, where individual photons strongly interact with each
other. As an example, we describe how these nonlinear processes
may be exploited to implement a single-photon transistor. While
ideas for developing plasmonic analogues of electronic devices by
combining SPs with electronics are already being
explored~\cite{atwater07}, the process we describe here opens up
fundamentally new possibilities, in that it combines the ideas of
plasmonics with the tools of quantum
optics~\cite{harris97,lukin03,fleischhauer05,miller05} to achieve
unprecedented control over the interactions of individual light
quanta.

\section{Nanowire plasmons: interaction with matter}

\begin{figure*}[p]
\begin{center}
\includegraphics[width=13cm]{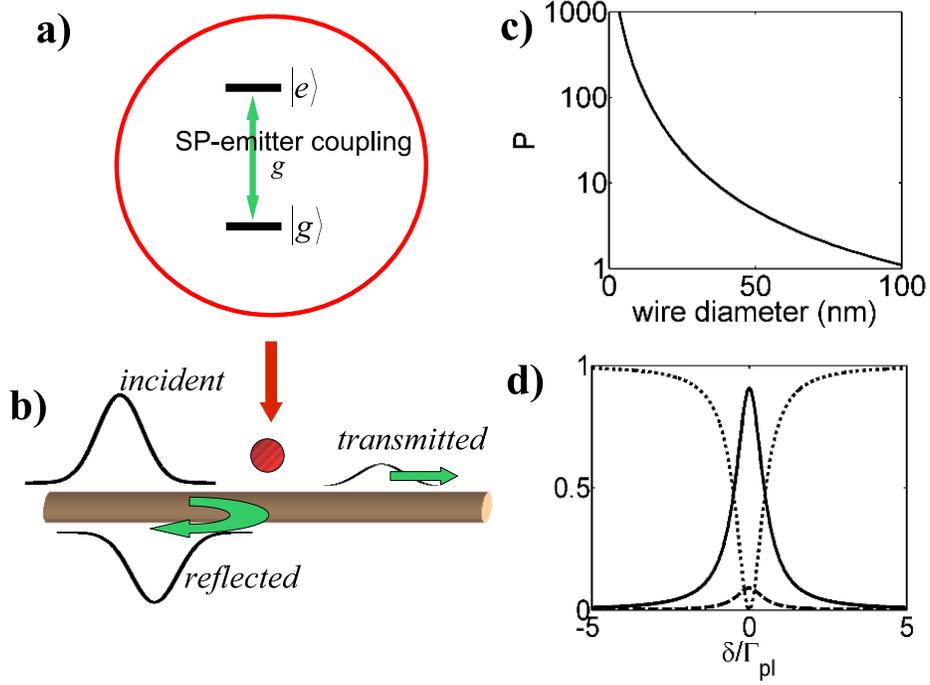}
\end{center}
\caption{a) Two-level emitter
interacting with the nanowire. States $\ket{g},\ket{e}$ are
coupled via the SP modes with a strength $g$. b) Schematic of a
single incident photon scattered off of a near-resonant emitter.
The interaction leads to reflected and transmitted fields whose
amplitudes can be calculated exactly. c) The maximum Purcell factor of an emitter positioned
near a silver nanowire~($\epsilon{\approx}-50+0.6i$) and
surrounded by uniform dielectric~($\epsilon=2$), as a function of
wire diameter. The plot is calculated using the method of Refs.~\cite{chang06a,chang06b} and the silver
properties used correspond to a free-space
wavelength of $\lambda_0=1\,\mu$m. d) Probabilities of
reflection~(solid line), transmission~(dotted line), and
loss~(dashed line) for a single photon incident upon a single
emitter, as a function of detuning. The Purcell factor for this
system is taken to be $P=20$. \label{fig:threelevels}}
\end{figure*}

SPs are propagating electromagnetic modes confined to the surface
of a conductor-dielectric interface~\cite{maier06,atwater07}.
Their unique properties make it possible to confine them to
sub-wavelength dimensions, which has led to fascinating new
approaches to waveguiding below the diffraction
limit~\cite{takahara97}, enhanced transmission through
sub-wavelength apertures~\cite{ebbesen07}, and sub-wavelength
imaging~\cite{letokhov02,smolyaninov05,zayats05}. Large field
enhancements associated with plasmon resonances of metallic
nano-particles have also been utilized to detect nearby single
molecules via surface-enhanced Raman
scattering~\cite{kneipp97,nie97}. Similar properties directly give
rise to the strong interaction between single SPs on a conducting
nanowire and an individual, proximal optical emitter~(see Figs.~\ref{fig:threelevels}a,~b), as we
describe below.

Much like in a single-mode fiber, the SP modes of a conducting
nanowire constitute a one-dimensional, single-mode continuum that
can be indexed by the wavevectors $k$ along the direction of
propagation~\cite{takahara97,chang06a,chang06b}. Unlike a
single-mode fiber~\cite{tong04}, however, the nanowire continues
to display good confinement and guiding when its radius is reduced
well below the optical wavelength~($R{\ll}\lambda_0$).
Specifically, in this limit, the SPs exhibit strongly reduced
wavelengths and small transverse mode areas relative to free-space
radiation, which scale like
$\lambda_{\footnotesize\textrm{pl}}{\propto}1/k{\propto}R$ and
$\Aeff{\propto}R^2$, respectively. The tight confinement results
in a large interaction strength between the SP modes and any
proximal emitter with a dipole-allowed transition, with a coupling
constant that scales like $g\propto{1}/\sqrt{\Aeff}$. The
reduction in group velocity also yields an enhancement of the
density of states, $D(\omega)\propto1/R$. The spontaneous emission
rate into the SPs,
$\Gammapl{\sim}g^2(\omega)D(\omega)\propto(\lambda_0/R)^3$, can
therefore be much larger than the emission rate $\Gamma'$ into all
other possible channels.  A relevant figure of merit is an
effective Purcell factor $P\equiv\Gammapl/\Gamma'$, which can
exceed $10^3$ in realistic systems~(see
Fig.~\ref{fig:threelevels}c). As will be seen, the Purcell factor
plays an important role in determining the strength and fidelity
of the nonlinear processes of interest.

Motivated by these considerations, we now describe a general
one-dimensional model of an emitter strongly coupled to a set of
travelling electromagnetic modes~(see
Figs.~\ref{fig:threelevels}a,~b). We first consider a simple two-level
configuration for the emitter, consisting of ground and excited
states $\ket{g},\ket{e}$ separated by frequency $\omega_{eg}$. The
Hamiltonian describing this system is given by
\be H= \hbar(\omega_{eg}-i\Gamma'/2)\sigma_{ee}
+\int\,dk\,\hbar{c}|k|\opdagger{a}{k}\op{a}{k}-\hbar{g}\int\,dk\,\left(\sigma_{eg}\op{a}{k}e^{ikz_a}+h.c.\right),\label{eq:Hfull}
\ee
where $\sigma_{ij}=\ket{i}\bra{j}$, $\op{a}{k}$ is the
annihilation operator for the mode with wavevector $k$, $g$ is the
emitter-field interaction matrix element, and $z_a$ is the
position of the emitter. We have assumed that a linear dispersion
relation holds over the relevant frequency range, $\nu_k=c|k|$,
where $c$ is the group velocity of SPs on the nanowire, and
similarly that $g$ is frequency-independent. In the spirit of the
stochastic wave function or ``quantum jump'' description of an
open system~\cite{meystre99}, we have also included a
non-Hermitian term in $H$ due to the decay of state $\ket{e}$ into
a reservoir of other radiative and non-radiative modes at a rate
$\Gamma'$.

\section{Single emitter as a saturable mirror}

The propagation of SPs can be dramatically altered by interaction
with the single two-level emitter. In particular, for low incident
powers, the interaction occurs with near-unit probability, and
each photon can be reflected with very high efficiency.  At the
same time, for higher powers the emitter response rapidly
saturates, as it is not able to scatter more than one photon at a
time.

The low-power behavior can be easily understood by first
considering the scattering of a single photon, as illustrated
schematically in Fig.~\ref{fig:threelevels}b. Since we are
interested only in SP modes near the optical frequency
$\omega_{eg}$, we can effectively treat left- and
right-propagating SPs as completely separate fields.  In
particular, one can define operators that annihilate a
left~(right)-propagating photon at position $z$,
$\op{E}{L(R)}(z)=(1/\sqrt{2\pi})\int\,dk\,e^{ikz}\op{a}{L(R),k}$,
where operators acting on the left and right branches are assumed
to have vanishing commutation relations with the other branch. An
exact solution to the scattering from the right to left branches
in the limit $P\rightarrow\infty$ was derived in~\cite{fan05}, and
this approach can be generalized to finite $P$.  In particular, it
is possible to solve for the scattering eigenstates of a system
containing at most one~(either atomic or photonic) excitation, as
described in Methods. The reflection coefficient for an incoming
photon of wavevector $k$ is
\be
r(\delta_k)=-\frac{1}{1+\Gamma'/\Gammapl-2i\delta_k/\Gammapl},\label{eq:rwire}
\ee
where $\delta_k{\equiv}ck-\omega_{eg}$ is the photon detuning,
while the transmission coefficient is related to $r$ by
$t(\delta_k)=1+r(\delta_k)$. Here $\Gammapl=4\pi{g^2}/c$ is the
decay rate into the SPs, as obtained by application of Fermi's
Golden Rule to the Hamiltonian in Eq.~(\ref{eq:Hfull}). On
resonance, $r{\approx}-(1-1/P)$, and thus for large Purcell
factors the emitter in state $\ket{g}$ acts as a nearly perfect
mirror, which simultaneously imparts a $\pi$-phase shift upon
reflection. The bandwidth $\Delta\omega$ of this process is
determined by the total spontaneous emission rate,
$\Gamma=\Gammapl+\Gamma'$, which can be quite large. Furthermore,
the probability $\kappa$ of losing the photon to the environment
is strongly suppressed for large Purcell factors,
$\kappa{\equiv}1-\mathcal{R}-\mathcal{T}=2\mathcal{R}/P$, where
$\mathcal{R}\,(\mathcal{T})\equiv|r|^2\,(|t|^2)$ is the
reflectance~(transmittance). These results are illustrated in
Fig.~\ref{fig:threelevels}d, where
$\mathcal{R},\mathcal{T},\kappa$ are plotted as a function of
detuning $\delta_k$, taking a conservative value of $P=20$.

The nonlinear response of the system can be seen by considering
the interaction of a single emitter not just with a single photon,
but with multi-photon input states. While the reflectance and
transmittance for this system at low powers resemble those derived
for a single photon, the arrival of several photons within the
bandwidth $\Delta\omega{\sim}\Gamma$ saturates the atomic response
and these photons cannot be efficiently reflected.  To be
specific, we consider the case when the incident field consists of
a coherent state, the quantum mechanical state that most closely
corresponds to a classical field~\cite{meystre99}. We now describe
a mapping that allows the scattering dynamics to be solved
exactly. We assume that the incident field propagates to the right
and that the emitter is initially in the ground state, such that
the initial wave function can be written in the form
$\ket{\tilde{\psi}(t{\rightarrow}-\infty)}=D(\{\alpha_{k}e^{-i\nu_{k}t}\})\ket{vac}\ket{g}$,
where the displacement operator
$D(\{\alpha_k\})\equiv\exp(\int\,dk\,\opdagger{a}{R,k}\alpha_k-\alpha_k^{\ast}\op{a}{R,k})$~\cite{meystre99}
creates a multi-mode coherent state from vacuum. This property of
the displacement operator motivates a state transformation given
by~\cite{mollow75}
\be
\ket{\tilde{\psi}}=D\left(\{\alpha_{k}e^{-i\nu_k{t}}\}\right)\ket{\psi},\label{eq:transformation}
\ee
so that the initial state is transformed into
$\ket{\psi(t{\rightarrow}-\infty)}=\ket{vac}\ket{g}$. In the
Heisenberg picture, the field operator transforms as
$\op{E}{R}(z,t){\rightarrow}\op{E}{R}(z,t)+\mathcal{E}_{c}(z,t)$,
where the external field amplitude is
$\mathcal{E}_{c}(z,t)=(1/\sqrt{2\pi})\int\,dk\,\alpha_{k}e^{ikz-i\nu_{k}t}$.
The transformation thus maps the initial coherent state to a
$c$-number in the interaction Hamiltonian, which physically
corresponds to a classical Rabi frequency~(given by
$\Omega_c=\sqrt{2\pi}g\mathcal{E}_c$), while simultaneously
mapping the initial photonic state to vacuum. An important
consequence is that the dynamics of the emitter interacting with
the field modes can now be treated under the Wigner-Weisskopf
approximation, \textit{i.e.}, interaction with the vacuum modes
gives rise to an exponential decay rate from state $\ket{e}$ to
$\ket{g}$ at a rate $\Gamma$. The evolution of the atomic
operators consequently reduces to the usual Langevin-Bloch
equations~\cite{meystre99}, which enables all properties of the
atomic operators and the scattered field to be calculated~(see
Methods). For a narrow bandwidth~($\delta\omega{\ll}\Gamma$),
resonant~($\delta_k=0$) input field, the steady-state
transmittance and reflectance are found to be
\bea \mathcal{T} & = &
\frac{1+8(1+P)^2(\Omega_c/\Gamma)^2}{(1+P)^2(1+8(\Omega_c/\Gamma)^2)},
\\ \mathcal{R} & = &
\left(1+\frac{1}{P}\right)^{-2}\frac{1}{1+8(\Omega_c/\Gamma)^2}.
\eea
At low powers~($\Omega_c/\Gamma{\ll}1$), the emitter has
scattering properties identical to the single-photon case,
$\mathcal{R}{\approx}(1+1/P)^{-2},\mathcal{T}{\approx}(1+P)^{-2}$,
and for large Purcell factors the single emitter again acts as a
perfect mirror.  At high incident
powers~($\Omega_{c}/\Gamma{\gg}1$), however, the single emitter
saturates and most of the incoming photons are simply transmitted
past with no effect,
$\mathcal{T}{\rightarrow}1,\mathcal{R}{\sim}\mathcal{O}((\Gamma/\Omega_c)^2)$.
The significance of these results can be understood by noting that
saturation is achieved at a Rabi frequency $\Omega_{c}\sim\Gamma$
that, in the limit of large $P$, corresponds to a switching energy
of a single quantum~($\sim\hbar\nu$) within a pulse of duration
$\sim{1}/\Gamma$.

\section{Photon correlations}

The strongly nonlinear atomic response at the single-photon level
leads to dramatic modification of photon statistics that cannot be
captured by only considering average intensities. We now consider
higher-order correlations of the transmitted and reflected fields.
Specifically, we focus on the normalized second-order correlation
function for the outgoing field, $g^{(2)}_{R,L}(t)$, which for a
stationary process is defined as
\be
g^{(2)}_{\beta=R,L}(z,t)\equiv\avg{\opdagger{E}{\beta}(z,\tau)\opdagger{E}{\beta}(z,\tau+t)\op{E}{\beta}(z,\tau+t)\op{E}{\beta}(z,\tau)}/\avg{\opdagger{E}{\beta}(z,\tau)\op{E}{\beta}(z,\tau)}^2,
\ee
where $t$ denotes the difference between the two observation times
$\tau$ and $\tau+t$.

\begin{figure*}[p]
\begin{center}
\includegraphics[width=17cm]{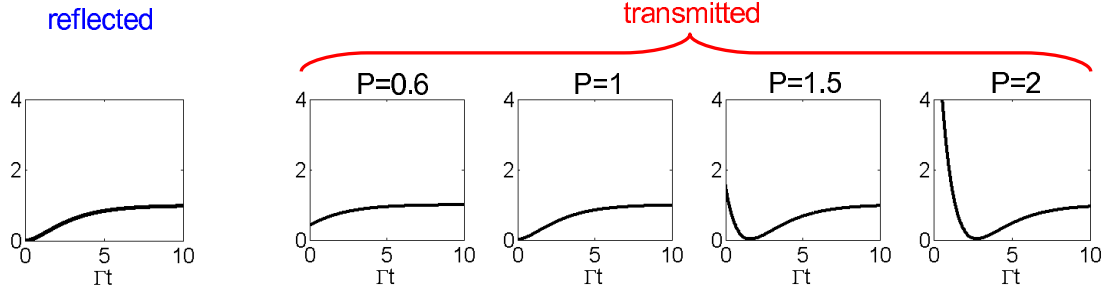}
\end{center}
\caption{Second-order correlation function $g^{(2)}(t)$ for the
reflected and transmitted fields at low incident
power~($\Omega_c/\Gamma=0.01$). $g^{(2)}(t)$ for the reflected
field is independent of $P$ at low powers.  For the transmitted
field, going from left to right, the Purcell factors are
$P=0.6,1,1.5,2$, respectively. A rise in $g^{(2)}(0)$ for large
Purcell factors indicates a strong initial bunching of photons at
the transmitted end.  This initial bunching is accompanied by an
anti-bunching effect, $g^{(2)}(t_0){\approx}0$, at some later time
$t_0=(4\log\,P)/\Gamma$ for $P{\geq}1$.  For high incident
powers~(not shown), $g^{(2)}(t)$ approaches unity for all times
due to a saturation of the atomic
response.\label{fig:twolevelcorrelations}}
\end{figure*}

The statistics of the reflected field is identical to the
well-known result for resonance fluorescence~\cite{meystre99} in
three dimensions~(see Fig.~\ref{fig:twolevelcorrelations}).  This
can intuitively be understood because it is a purely scattered
field.  It follows that the field is strongly anti-bunched,
$g^{(2)}(0)=0$, since the emitter can only absorb and re-emit one
photon at a time. The transmitted field, however, has unique
properties because it is a sum of the incident and scattered
fields. For near-resonant excitation, we find for low powers
that~(see Methods)
\be
g^{(2)}(t)=e^{-\Gamma{t}}\left(P^2-e^{\Gamma{t}/2}\right)^2+\mathcal{O}(\Omega_c^2/\Gamma^2),\label{eq:g2R}\ee
while for high powers $g^{(2)}(t)$ approaches unity for all times.
The high power result indicates that no change in statistics
occurs and is due to saturation of the atomic response. The
low-power behavior reflects that of an efficient single-photon
switch.  In particular, for $P{\gg}1$, individual photons have a
large reflection probability, but when two photons are incident
simultaneously the transition saturates, so that photon pairs have
a much larger probability of transmission~(for $P{\ll}1$ the
emitter has little influence and the statistics of the transmitted
field are almost unchanged). This phenomenon yields a strong
bunching effect at $t=0$ that behaves like
$g^{(2)}(0){\approx}P^4$.  One also finds a subsequent
anti-bunching and perfect vanishing of $g^{(2)}(t)$ at the later
time $t_0=(4\log\,P)/\Gamma$ for weak input fields.  A more
detailed understanding of these features can be gained from a
quantum jump picture describing the system's
evolution following the detection of a photon~\cite{carmichael91}. Unlike for the
reflected field, the picture for the transmitted field is more
complicated because one cannot determine whether the detected
photon originates from the emitter or from direct transmission of
the incident field. More formally, the change in the wave function
following detection is described by the application of a jump
operator to the system, given in this case by the transmitted
field operator,
$\op{E}{T}=\op{E}{R,free}+\mathcal{E}_{c}+\sqrt{2\pi}ig\sigma_{ge}/c$~(cf.
Eq.~(\ref{eq:transmittedfield}) in Methods). For large $P$,
$\op{E}{T}$ is strongly influenced by its atomic component.  This
is responsible for, \textit{e.g.}, the low transmittance
$\mathcal{T}{\approx}(1+P)^{-2}$ in steady-state, as the field
scattered by the emitter destructively interferes with the
incoming field.  Because multiple incident photons increase the
transmission probability, the detection of a photon enhances the
conditional probability that another photon is present in the
system.  In the quantum jump picture this translates into a sudden
enhancement of the coherence $\avg{\sigma_{ge}}$ by a factor of
$1+P$ over its steady-state value~(see Methods). The destructive
interference between the incoming and scattered fields is
subsequently lost, and the jump causes a sudden enhancement in the
field amplitude $\avg{\op{E}{T}}$ while also inducing a
$\pi$-phase shift relative to its equilibrium value. The initial
enhancement in $\avg{\op{E}{T}}$ gives rise to bunching.  Then,
the $\pi$-phase shift and subsequent relaxation back to
equilibrium causes $\avg{\op{E}{T}}$ to pass through zero at time
$t_0$, which yields the subsequent anti-bunching and reflects the
cancellation of the incoming and scattered fields. For $P=1$, this
cancellation happens exactly at $t=0$ such that $g^{(2)}(0)=0$.

\section{Ideal single-photon transistor}

\begin{figure*}[p]
\begin{center}
\includegraphics[width=17cm]{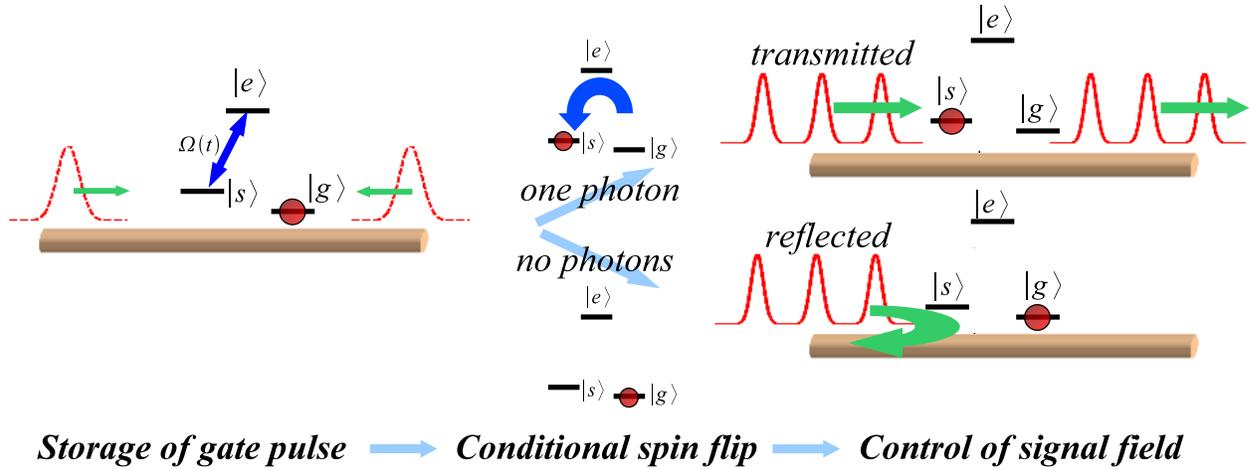}
\end{center}
\caption{Schematic of transistor operation involving a three-level
emitter. In the storage step, a gate pulse consisting of zero or
one photon is split equally in counter-propagating directions and
coherently stored using an impedance-matched control field
$\Omega(t)$. The storage results in a spin flip conditioned on the
photon number. A subsequent incident signal field is either
transmitted or reflected depending on the photon number of the
gate pulse, due to the sensitivity of the propagation to the
internal state of the emitter.\label{fig:transistor}}
\end{figure*}

While the two-level emitter analyzed previously is capable of
acting as a switch that distinguishes between single- and
multi-photon fields, a greater degree of coherent control can be
gained by considering the interaction of light with a multi-level
emitter. For concreteness, we consider the three-level
configuration shown in Fig.~\ref{fig:transistor}. Here, a
metastable state $\ket{s}$ is decoupled from the SPs due to,
\textit{e.g.}, a different orientation of its associated dipole
moment, but is resonantly coupled to $\ket{e}$ via some classical,
optical control field with Rabi frequency $\Omega(t)$. States
$\ket{g},\ket{e}$ remain coupled via the SP modes as discussed
earlier.  Using this system, we now describe a process in which a
single ``gate'' photon can completely control the propagation of
subsequent ``signal'' pulses consisting of either individual or
multiple photons, whose timing can be arbitrary. In analogy to the
electronic counterpart, this corresponds to an ideal single-photon
transistor.

We first describe how one can achieve coherent storage of a single
photon, and then how this can be combined with the reflective
properties derived above to realize a single-photon transistor.
The storage is an important ingredient, as it provides an atomic
memory of the gate field and hence allows the gate to interact
with the subsequent signal field. The idea behind single-photon
storage is to initialize the emitter in $\ket{g}$ and to apply the
control field $\Omega(t)$ simultaneous with the arrival of a
single photon in the SP modes. The control field, if properly
chosen~(or ``impedance-matched'')~\cite{cirac97}, will result in
capture of the incoming single photon while inducing a spin flip
from $\ket{g}$ to $\ket{s}$. Generally, one can show by time
reversal symmetry~\cite{gorshkov06} that the optimal storage
strategy is the time-reversed process of single-photon generation,
where the emitter is driven from $\ket{s}$ to $\ket{g}$ by the
external field while emitting a single photon, whose wavepacket
depends on $\Omega(t)$. By this argument, it is evident that
optimal storage is obtained by splitting the incoming pulse and
having it incident from both sides of the emitter
simultaneously~(see Fig.~\ref{fig:transistor}), and that there is
a one-to-one correspondence between the incoming pulse shape and
the optimal field $\Omega(t)$. Moreover, the storage efficiency is
identical to that of single photon generation and is thus given by
${\sim}1-1/P$ for large $P$~\cite{chang06a}.  This result is also
derived explicitly in Supplemental Information, where we solve for
the dynamics of this three-level system exactly. Physically, the
fidelity of storage is simply determined by the degree to which
coupling of the emitter to the SP modes exceeds the coupling to
other channels.  A detailed analysis reveals that this optimum is
achievable for any input pulse of duration $T{\gg}1/\Gamma$ and
for a certain class of pulses of duration $T\sim
1/\Gamma$~\cite{gorshkov06}. Finally, we note that if no photon
impinges upon the emitter, the pulse $\Omega(t)$ has no effect and
the emitter remains in state $\ket{g}$ for the entire process. The
result is more generally described as a mapping between single SP
states and metastable atomic states,
$(\alpha\ket{0}+\beta\ket{1})\ket{g}\rightarrow\ket{0}(\alpha\ket{g}+\beta\ket{s})$.

We next consider the reflection properties of the emitter when the
control field $\Omega(t)$ is turned off. If the emitter is in
$\ket{g}$, the reflectance and transmittance derived above for the
two-level emitter remain valid.  On the other hand, if the emitter
is in $\ket{s}$, any incident fields will simply be transmitted
with no effect since this state is decoupled from the SPs.
Therefore, with $\Omega(t)$ turned off, the three-level system
effectively behaves as a conditional mirror whose properties
depend sensitively on its internal state.

The techniques of state-dependent conditional reflection and
single-photon storage can be combined to create a single-photon
transistor, whose operation is illustrated in
Fig.~\ref{fig:transistor}. The key principle is to utilize the
presence or absence of a photon in an initial ``gate'' pulse to
conditionally flip the internal state of the emitter during the
storage process, and to then use this conditional flip to control
the flow of subsequent ``signal'' photons arriving at the emitter.
The first step is to implement the storage protocol for the gate
pulse, consisting of either zero or one photon, starting with the
emitter in $\ket{g}$. The presence~(absence) of a photon causes
the emitter to flip to~(remain in) state $\ket{s}$~($\ket{g}$).
Next, the interaction of each signal pulse arriving at the emitter
depends sensitively on the internal state that results after
storage.  The storage step and conditional spin flip causes the
emitter to be either highly reflecting or completely transparent
depending on the gate, and the system therefore acts as an
efficient switch or transistor for the subsequent signal field.

The ideal operation of the transistor is limited only by the
characteristic time over which an undesired spin flip can occur.
In particular, if the emitter remains in $\ket{g}$ after storage
of the gate pulse, the emitter can eventually be optically pumped
to $\ket{s}$ upon the arrival of a sufficiently large number of
photons in the signal field. For strong coupling the number of
incident photons $n$ that can be scattered before pumping occurs
is given by the branching ratio of decay rates from $\ket{e}$ to
these states,
$n{\sim}\Gamma_{e{\rightarrow}g}/\Gamma_{e{\rightarrow}s}$, which
can be large due to the large decay rate
$\Gamma_{e\rightarrow{g}}\geq\Gammapl$.  Thus $n{\gtrsim}P$ and
the emitter can reflect $\mathcal{O}(P)$ photons before an
undesired spin flip occurs. This number corresponds to the
effective ``gain'' of the single-photon transistor.

Finally, we note that there exist other possible realizations of a
single-photon transistor as well.
The ``impedance-matching'' condition and the need to split a pulse
for optimal storage, for example, can be relaxed using a small
ensemble of emitters and photon storage techniques based on
electromagnetically induced
transparency~(EIT)~\cite{fleischhauer00}. Here, storage also
results in a spin flip within the ensemble that sensitively alters
the propagation of subsequent photons.

\section{Integrated systems}

\begin{figure*}[p]
\begin{center}
\includegraphics[width=15cm]{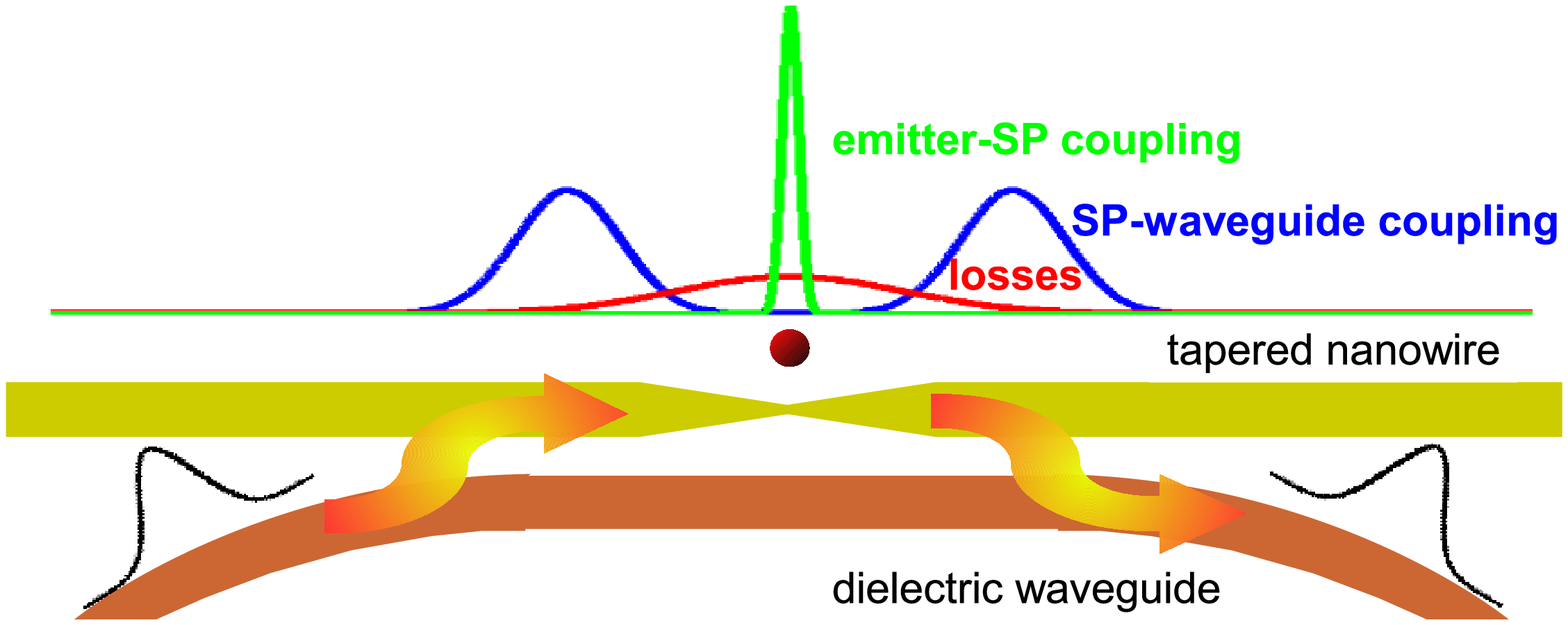}
\end{center}
\caption{Illustration of in-and out-coupling of SPs on a tapered
nanowire to an evanescently coupled, low-loss dielectric
waveguide. Here, a single photon originally in the waveguide is
transferred to the nanowire, where it interacts with the emitter
before being transferred back into the waveguide.  The coupling
between the nanowire and waveguide is efficient only when they are
phase-matched~(in the regions indicated by the blue peaks). The
phase-matching condition is poor in the regions of the wire taper
and in the bending region of the waveguide away from the nanowire.
Dissipative losses~(in red) are concentrated to a small region
near the nanowire taper, due to a large concentration of fields
here.\label{fig:outcoupling}}
\end{figure*}

Thus far we have not dealt with the inevitable losses that SPs
experience as they propagate along the nanowire, which could
potentially limit their feasibility as long-distance carriers of
information and their use in large-scale devices. For the
nanowire, one must consider the trade-off between the larger
Purcell factors obtainable with smaller diameters and a
commensurate increase in dissipation due to the tighter field
confinement. However, these limitations are not fundamental, if
one can integrate SP devices with low-loss dielectric waveguides
and other microphotonic devices.  Here, the SPs can be used to
achieve strong nonlinear interactions over very short interaction
distances, but are rapidly in- and out-coupled to conventional
waveguides for long-distance transport.  One such integration
scheme is illustrated in Fig.~\ref{fig:outcoupling}, where
excitations are transferred to and from the SP modes of a nanowire
from an evanescently coupled, phase-matched dielectric waveguide.
The losses will be small provided that the distance needed for the
SPs to be coupled in and out and interact with the emitter is
smaller than the characteristic dissipation length. This can be
accomplished by techniques such as optimizing of SP
geometries~(\textit{e.g.}, tapered wires or
nanotips~\cite{chang06a,chang06b}) and engineering of SP
dispersion relations~\cite{maier05} via periodic structures. Coupling efficiencies exceeding 95\%, for example, are predicted using simple systems~\cite{chang06b}. Such
a conductor/dielectric interface would provide convenient
integration with conventional optical elements, enable many
nonlinear operations without loss, and make large-scale,
integrated photonic devices feasible.

Another key feature of nano-scale SPs is that the strong
interaction with a single emitter is very robust.  In particular,
the large coupling occurs over a very large bandwidth and no
special tuning of either the emitter or nanowire is required.  SPs
are thus promising candidates for use with solid-state emitters
such as quantum dot nanocrystals~\cite{klimov00} or
color centers~\cite{brouri00}, where the spectral properties can
vary over individual emitters. Color centers in
diamond~\cite{tamarat06}, for instance, are especially promising
because they offer sharp optical lines and three-level internal
configurations. At the same time, guided SPs might be used for
trapping isolated neutral atoms in the vicinity of suspended
wires, thereby creating an effective interface for isolated atomic
systems.

\section{Outlook}

We now outline some new directions opened up by this work. We have
shown that a single emitter near a conducting nanowire provides a
strong optical nonlinearity at the level of single photons, which
can be exploited to create a single-photon transistor. This can be
used for a variety of important applications, such as very
efficient single-photon detection, where the large gain in the
signal field provides for efficient detection of the gate pulse.
This system also finds applications in quantum information
science. One can prepare Schrodinger cat states of photons, for
example, if the gate pulse contains a superposition of zero and
one photon, since this initial pulse can be entangled with the
propagation direction of a large number of subsequent signal
photons. The controlled-phase gate for photons proposed
in~\cite{duan04} for cavity QED is also directly extendable to our
plasmonic system. In particular, this scheme relies on the
conditional phase shifts acquired as photons are reflected from a
resonant cavity containing a single atom, which are analogous to
the reflection dynamics derived for single SPs here. In addition,
by using SPs it is possible to achieve very large optical depths
with just a few emitters, which would make this system effective
for realizing EIT-based nonlinear schemes~\cite{lukin00}.
Furthermore, the present system is an intriguing candidate to
observe phenomena associated with strongly interacting,
one-dimensional many-body systems. For example, non-perturbative
effects such as photon-atom bound states~\cite{leclair97} and
quantum phase transitions~\cite{lesage98} involving photons can be
explored. Higher-order correlations created in the transmitted
field can become a useful tool to study and probe the
non-equilibrium quantum dynamics of these strongly interacting
photonic systems.

\section{Methods}

\subsection{Single-photon dynamics}

Because we are interested only in the dynamics of near-resonant
photons with an emitter, we can make the approximation that left-
and right-propagating photons form completely separate quantum
fields~\cite{fan05}. We define annihilation and creation operators
for the two fields, $\op{a}{L(R),k},\opdagger{a}{L(R),k}$, where
the index $k$ is assumed to run over the range $\pm\infty$; in
principle this allows for the existence of negative-energy modes,
but this is unimportant if we consider near-resonant dynamics.
Under this two-branch approximation, the relevant terms in
Eq.~(\ref{eq:Hfull}) are transformed via
$\int\,dk\,\hbar{c}|k|\opdagger{a}{k}\op{a}{k}\rightarrow\int\,dk\,\hbar{c}k\left(\opdagger{a}{R,k}\op{a}{R,k}+\opdagger{a}{L,-k}\op{a}{L,-k}\right)$
and
$\sigma_{eg}\op{a}{k}e^{ikz_a}\rightarrow\sigma_{eg}\left(\op{a}{R,k}+\op{a}{L,k}\right)e^{ikz_a}$.

To solve for the reflection and transmission coefficients of
single-photon scattering, we write the general wave function for a
system containing one~(either photonic or atomic) excitation in
the following way~(here a two-level emitter is assumed),
\be
\ket{\psi_k}=\int\,dz\left(\phi_{L}(z)\opdagger{E}{L}(z)+\phi_{R}(z)\opdagger{E}{R}(z)\right)\ket{g,vac}+c_{e}\ket{e,vac}.
\ee
The field amplitudes are chosen to correspond to photons of
well-defined momenta in the limits $z{\rightarrow}\pm\infty$,
\textit{e.g.}, $\phi_{R}(z{\rightarrow}-\infty){\sim}e^{ikz}$,
$\phi_{R}(z{\rightarrow}\infty){\sim}te^{ikz}$, and
$\phi_{L}(z\rightarrow{-\infty})\sim{r}e^{-ikz}$ for a photon
propagating initially to the right, where $t$~($r$) is the
transmission~(reflection) coefficient. Following~\cite{fan05}, we
obtain Eq.~(\ref{eq:rwire}) by solving the time-independent
Schrodinger equation $H\ket{\psi_k}=E_k\ket{\psi_k}$ for $r,t$ and
$c_e$.

\subsection{Multi-photon dynamics}

In the two-branch approximation, the Heisenberg equations of
motion for the fields are given by
\be
\left(\frac{\partial}{{\partial}z}+\frac{1}{c}\frac{\partial}{{\partial}t}\right)\op{E}{R}(z,t)=
\frac{\sqrt{2\pi}ig}{c}\sigma_{ge}(t)\delta(z-z_a), \ee
which can be formally integrated to give
\be \op{E}{R}(z,t)=
\op{E}{R,free}(z-ct)+\frac{\sqrt{2\pi}ig}{c}\sigma_{ge}\left(t-(z-z_a)/c\right)\Theta(z-z_a)\label{eq:transmittedfield},
\ee
where $\Theta(z)$ is the Heaviside step function.  A similar
equation holds for $\op{E}{L}$. Assuming that the field initially
propagates to the right, $\op{E}{R}(z,t)$ is the field transmitted
past the emitter for $z>z_a$, while for $z<z_a$, $\op{E}{L}(z,t)$
is the reflected field.

We now discuss how to calculate the transmitted field intensity~(a
similar method holds for finding the reflected intensity). Under
the transformation given by Eq.~(\ref{eq:transformation}), the
first-order correlation function for the right-going field is
given by
\be
G^{(1)}_{R}(z,t)=\avg{(\opdagger{E}{R}(z,t)+\mathcal{E}_{c}^{\ast}(z,t))(\op{E}{R}(z,t)+\mathcal{E}_{c}(z,t))},\label{eq:G1}
\ee
which upon evaluating at $z>z_a$ yields the average transmitted
intensity. We proceed by substituting
Eq.~(\ref{eq:transmittedfield}) into Eq.~(\ref{eq:G1}). Because
the initial photonic state is vacuum following the transformation,
$\op{E}{R,free}$ has no effect and thus calculation of $G^{(1)}$
reduces to calculating correlations between atomic operators.
Techniques for evaluating these correlations are well-known using
the Langevin-Bloch equations~\cite{meystre99}. Calculation of
$g^{(2)}(t)$ proceeds in a similar manner by using
Eq.~(\ref{eq:transmittedfield}) to express $g^{(2)}(t)$ in terms
of two-time atomic correlations, which can be evaluated using the
well-known quantum regression theorem~\cite{meystre99}.

The system in consideration undergoes a quantum jump following
detection of a transmitted photon.  Immediately following the
detection, the density matrix is given by
$\rho_{jump}=\op{E}{T}\rho_{ss}\opdagger{E}{T}/\avg{\opdagger{E}{T}\op{E}{T}}_{ss}$,
where $\rho_{ss}$ is the steady-state density matrix and
$\avg{{}}_{ss}$ denotes the average of quantities in steady state.
Here $\op{E}{T}$ is the jump operator defined in the ``Photon
correlations'' section and physically corresponds to the
transmitted field. In the weak-field limit, it is straightforward
to show that
$\avg{\sigma_{ge}}_{jump}=(1+P)\avg{\sigma_{ge}}_{ss}=2i\Omega_c/\Gamma'$,
and $\avg{\op{E}{T}}_{jump}/\avg{\op{E}{T}}_{ss}=1-P^2$.  Note in
particular that for large $P$, there is an initial enhancement in
the transmitted field amplitude as well as a $\pi$-phase shift
from its equilibrium value.


\begin{acknowledgements}
We thank A. Akimov, A. Mukherjee, V. Gritsev, M. Loncar, and H.
Park for useful discussions. This work was supported by the NSF (Career and NIRT programs), Harvard-MIT CUA, and Danish Natural Science Research
Council.
\end{acknowledgements}


\begin{thebibliography}{99}
\bibitem{boyd92} Boyd, R.W. \textit{Nonlinear Optics} (Academic
Press, New York, 1992).
\bibitem{gibbs85} Gibbs, H.M. \textit{Optical bistability: controlling light with light}
(Academic Press, Inc., Orlando, FL 1985).
\bibitem{bouwmeester00} Bouwmeester, D., Ekert, A. \& Zeilinger,
A., Eds. \textit{The Physics of Quantum Information} (Springer,
Berlin, 2000).
\bibitem{schmidt96} Schmidt, H. \& Imamoglu, A. Giant Kerr nonlinearities obtained by electromagnetically induced transparency. \textit{Opt. Lett.} {\bf 21}, 1936 (1996).
\bibitem{harris97} Harris, S.E. Electromagnetically Induced
Transparency. \textit{Phys. Today} {\bf 50}, 36 (1997).
\bibitem{harris98} Harris, S.E. \& Yamamoto, Y. Photon Switching
by Quantum Interference. \textit{Phys. Rev. Lett.} {\bf 81}, 3611
(1998).
\bibitem{lukin03} Lukin, M.D. Colloquium: Trapping and
manipulating photon states in atomc ensembles. \textit{Rev. Mod.
Phys.} \textbf{75}, 457 (2003).
\bibitem{fleischhauer05} Fleischhauer, M., Imamoglu, A. \&
Marangos, J.P. Electromagnetically induced transparency: Optics in
coherent media. \textit{Rev. Mod. Phys.} {\bf 77}, 633 (2005).
\bibitem{vahala04} Vahala, K., Ed. \textit{Optical Microcavities}
(World Scientific, Singapore, 2004).
\bibitem{miller05} Miller, R. \textit{et al}. Trapped atoms in cavity
QED: coupling quantized light and matter. \textit{J. Phys. B: At.
Mol. Opt. Phys.} {\bf 38}, S551 (2005).
\bibitem{duan04} Duan, L.-M. \& Kimble, H.J.  Scalable Photonic Quantum Computation through Cavity-Assisted Interactions. \textit{Phys. Rev Lett.} {\bf 92}, 127902
(2004).
\bibitem{birnbaum05} Birnbaum, K.M. \textit{et al.} Photon blockade in an optical cavity with one trapped atom. \textit{Nature} {\bf
436}, 87 (2005).
\bibitem{waks06} Waks, E. \& Vuckovic, J. Dipole Induced
Transparency in drop filter cavity-waveguide systems.
\textit{Phys. Rev. Lett.} {\bf 96}, 153601 (2006).
\bibitem{chang06a} Chang, D.E., S{\o}rensen, A.S., Hemmer, P.R. \&
Lukin, M.D. Quantum Optics with Surface Plasmons. \textit{Phys.
Rev. Lett.} {\bf 97}, 053002 (2006).
\bibitem{akimov07} Akimov, A.V. \textit{et al.}, submitted to Nature (2007).
\bibitem{atwater07} Atwater, H.A. The promise of
plasmonics. \textit{Scientific American} \textbf{296}, 56 (2007).
\bibitem{maier06} Maier, S.A. \textit{Plasmonics: fundamentals and
applications} (Springer-Verlag, New York, 2006).
\bibitem{takahara97} Takahara, J., Yamagishi, S., Taki, H., Morimoto, A. \& Kobayashi, T. Guiding of a one-dimensional optical beam with nanometer diameter. \textit{Opt. Lett.} {\bf
22},475 (1998).
\bibitem{ebbesen07} Genet, C. \& Ebbesen, T.W. Light in
tiny holes. \textit{Nature} {\bf 445}, 39 (2007).
\bibitem{letokhov02} Klimov, V.V., Ducloy, M. \& Letokhov, V.S. A
model of an apertureless scanning microscope with a prolate
nanospheriod as a tip and an excited molecule as an object.
\textit{Chem. Phys. Lett.} {\bf 358}, 192 (2002).
\bibitem{smolyaninov05} Smolyaninov, I.I., Elliott, J., Zayats, A.V. \& Davis, C.C. Far-Field Optical Microscopy with a Nanometer-Scale Resolution Based on the In-Plane Image Magnification by Surface Plasmon Polaritons. \textit{Phys. Rev. Lett.} {\bf
94}, 057401 (2005).
\bibitem{zayats05} Zayats, A.V., Elliott, J., Smolyaninov, I.I. \& Davis, C.C. Imaging with short-wavelength surface plasmon polaritons. \textit{Appl. Phys. Lett.} {\bf 86},
151114 (2005).
\bibitem{kneipp97} Kneipp, K. \textit {et al}. Single Molecule Detection Using Surface-Enhanced Raman Scattering (SERS). \textit{Phys. Rev. Lett}. {\bf 78}, 1667
(1997).
\bibitem{nie97} Nie, S. \& Emory, S.R. Probing Single Molecules and Single Nanoparticles by Surface-Enhanced Raman Scattering. \textit{Science} {\bf 275}, 1102 (1997).
\bibitem{chang06b} Chang, D.E., S{\o}rensen, A.S., Hemmer, P.R. \& Lukin,
M.D. Strong coupling of single emitters to surface plasmons.
quant-ph/0603221.
\bibitem{tong04} Tong, L., Lou, J. \& Mazur, E. Single-mode guiding properties of subwavelength-diameter silica and silicon wire waveguides. \textit{Opt. Express} {\bf 12}, 1025
(2004).
\bibitem{meystre99} Meystre, P. \& Sargent III, M.
\textit{Elements of Quantum Optics}, 3rd ed. (Springer-Verlag, New
York, 1999).
\bibitem{fan05} Shen, J.T. \& Fan, S. Coherent photon transport from spontaneous emission in one-dimensional waveguides. \textit{Opt. Lett.} {\bf 30}, 2001 (2005).
\bibitem{mollow75} Mollow, B.R. Pure-state analysis of resonant light scattering: Radiative damping, saturation, and multiphoton effects. \textit{Phys. Rev. A} {\bf 12}, 1919
(1975).
\bibitem{carmichael91} Carmichael, H.J., Brecha, R.J. \& Rice, P.R. Quantum interference and collapse of the wavefunction in cavity QED. \textit{Opt. Comm.} {\bf 82}, 73 (1991).
\bibitem{cirac97} Cirac, J.I., Zoller, P., Kimble, H.J. \& Mabuchi, M. Quantum State Transfer and Entanglement Distribution among Distant Nodes in a Quantum Network. \textit{Phys. Rev. Lett}.
{\bf 78}, 3221 (1997).
\bibitem{gorshkov06} Gorshkov, A.V., Andre, A., Fleischhauer, M., S{\o}rensen, A.S. \& Lukin, M.D. Universal Approach to Optimal Photon Storage in Atomic Media. \textit{Phys. Rev. Lett}. {\bf 98}, 123601 (2007).
\bibitem{fleischhauer00} Fleischhauer, M. \& Lukin, M.D. Dark-State Polaritons in Electromagnetically Induced Transparency. \textit{Phys. Rev. Lett.} {\bf
84}, 5094 (2000).
\bibitem{maier05} Maier, S.A., Friedman, M.D., Barclay, P.E. \& Painter, O. Experimental demonstration of fiber-accessible metal nanoparticle plasmon waveguides for planar energy guiding and sensing. \textit{Appl. Phys.
Lett.}
{\bf 86}, 071103 (2005).
\bibitem{klimov00} Klimov, V.I. \textit{et al}. Optical Gain and Stimulated Emission in Nanocrystal Quantum Dots. \textit{Science} {\bf 290},
314 (2000).
\bibitem{brouri00} Brouri, R., Beveratos, A., Poizat, J.-P. \& Grangier, P. Photon antibunching in the fluorescence of individual color centers in diamond. \textit{Opt. Lett.} {\bf 25},
1294 (2000).
\bibitem{tamarat06} Tamarat, Ph. \textit{et al}. Stark Shift
Control of Single Optical Centers in Diamond. \textit{Phys. Rev.
Lett.} {\bf 97}, 083002 (2006).
\bibitem{lukin00} Lukin, M.D. \& Imamoglu, A. Nonlinear Optics and Quantum Entanglement of Ultraslow Single Photons. \textit{Phys. Rev. Lett.} {\bf 84}, 1419
(2000).
\bibitem{leclair97} Leclair, A., Lesage, F., Lukyanov, S. \& Saleur, H. The Maxwell-Bloch theory in quantum optics and the Kondo model. \textit{Phys. Lett. A} {\bf
235}, 203 (1997).
\bibitem{lesage98} Lesage, F. \& Saleur, H. Boundary Interaction Changing Operators and Dynamical Correlations in Quantum Impurity Problems. \textit{Phys. Rev. Lett.} {\bf
80}, 4370 (1998).
\end{thebibliography}
\end{document}